\documentclass[runningheads]{llncs}
\usepackage[T1]{fontenc}
\usepackage{cite}
\usepackage{amsmath,amssymb,amsfonts}
\usepackage{graphicx}
\usepackage{textcomp}
\usepackage{xcolor}
\usepackage{bm}
\usepackage{mathrsfs}
\usepackage{enumerate}
\usepackage{multirow}
\usepackage{color}
\usepackage{threeparttable}
\usepackage{subfigure}
\usepackage{booktabs}
\usepackage{times}
\usepackage{array}
\usepackage{verbatim}
\usepackage{algorithm}
\usepackage{bbding}
\usepackage{algpseudocode}
\usepackage{lettrine}
\usepackage{pifont}
\usepackage{float}
\newcommand{\cmark}{\ding{51}}%
\newcommand{\xmark}{\ding{55}}%
\begin{document}
\title{FreqU-FNet: Frequency-Aware U-Net for Imbalanced Medical Image Segmentation}
%
%
\author{Ruiqi Xing\inst{1}\orcidID{0009-0006-5139-8pap48}}
\authorrunning{Ruiqi Xing}
%
\institute{South China University of Technology, Guangdong,China
\email{202330333081@mail.scut.edu.cn}}
\maketitle              
\begin{abstract}
Medical image segmentation faces persistent challenges due to severe class imbalance and the frequency-specific distribution of anatomical structures. Most conventional CNN-based methods operate in the spatial domain and struggle to capture minority class signals, often affected by frequency aliasing and limited spectral selectivity. Transformer-based models, while powerful in modeling global dependencies, tend to overlook critical local details necessary for fine-grained segmentation. To overcome these limitations, we propose FreqU-FNet, a novel U-shaped segmentation architecture operating in the frequency domain. Our framework incorporates a Frequency Encoder that leverages Low-Pass Frequency Convolution and Daubechies wavelet-based downsampling to extract multi-scale spectral features. To reconstruct fine spatial details, we introduce a Spatial Learnable Decoder (SLD) equipped with an adaptive multi-branch upsampling strategy. Furthermore, we design a frequency-aware loss (FAL) function to enhance minority class learning. Extensive experiments on multiple medical segmentation benchmarks demonstrate that FreqU-FNet consistently outperforms both CNN and Transformer baselines, particularly in handling under-represented classes, by effectively exploiting discriminative frequency bands.

\end{abstract}
\section{Introduction}

Medical image segmentation is a fundamental task in clinical diagnosis, disease monitoring and treatment planning, playing a crucial role in precisely identifying anatomical structures. However, segmentation accuracy of multi-organ classes are heavily depends on capturing intricate anatomical details and multi-scale features from medical images, a challenge still inadequately addressed by existing methodologies, thus underscoring the motivation of this study.

Traditional CNN-based segmentation approaches primarily focus on extracting and processing features in spatial domain. Renowned models such as U-Net~\cite{ronneberger2015u} and UNet++\cite{8932614} rely primarily on convolution combined with traditional pooling and straightforward feature concatenation techniques. nnUNet\cite{isensee2021nnu} then improve the segmentation performance by introducing standardlized adaptive framework that automatically optimizes data preprocessing, model architecture an training protocols based on dataset characteristics. Although these models provide stable from spatial-frequency aliasing~\cite{chen2024semantic} due to simplistic pooling operations. Moreover, directly adding upsampled coarse features and high-resolution features from different levels resulting in blurred boundaries and category inconsistency inside the segmented objects.

Transformer-based architectures, exemplified by TransUNet~\cite{chen2021transunet} and Swin-UNet~\cite{cao2022swin} have emerged as powerful alternatives, effectively capturing global contextual relationships within medical images through self-attention. Yet these models are easily overfitted without extra datasets, which require to pretrained on large-scale annotated datasets. More critically, current transformer-based approaches failed to address the significant variations in segmentation accuracy across different anatomical structures persist since self-attention module focus more on global context information. Furthermore, empirical evidence\cite{isensee2024nnu} suggests that these architectures do not consistently surpass classical UNet structure in clinical segmentation tasks, particularly when evaluated under data-constrained scenarios common in medical imaging applications.Furthermore, most of upsample method consider single convolution as upsampling tool, overlooking the effect of multi-branch upsampling and point sampling which could be effective.

To address these critical issues, we introduce the Frequency domain Unet-like Fusion Network (FreqU-FNet), a novel architecture that innovatively operates in the frequency domain to effectively capture class-specific features across different frequency bands. Specifically, the encoder are based mainly on frequency domain, employing Frequency Low pass Convolution combined with Daubechies wavelet-based downsampling, substantially reducing aliasing artifacts while preserving critical anatomical information, particularly for minority classes that manifest in specific frequency ranges. The decoder is a Spatial Learnable Decoder (SLD) that integrates an Adaptive Hierarchical Multi-Branch Upsampling, helping the network to recover the spatial information. The method blends pixel-level details and spatial-contextual information through adaptive weighting of distinct upsampling pathways tailored to different frequency components.
To enable decoder extract semantic feature and rebuild the images on spatial domain, we introduce a spatial auxiliary learning module inspired by deep supervision ~\cite{wang2015training} to help the module capture basic spatial domain information while remaining the model to learn mostly on frequency domain. Correspondingly, we modify the loss function to instruct model segment based on frequency domain.
In summary, the contributions of our work are:
\begin{itemize}
    \item A new Frequency domain-based CNN segmentation network, efficiently using frequency domain information, reduced frequency aliasing, obtain signal frequency selectivity and preserve detailed information.
    \item A Spatial Learnable Decoder (SLD) that incorporates an adaptive hierarchical multi-branch upsampling component, collecting both pixel-level detail and broader spatial-contextual information through self-apative upsampling.
    \item A loss function, using both spatial and frequency domain information that could instruct model to learn on frequency domain and handle imbalance class problem.
\end{itemize}

\begin{figure*}[t]
    \centering
    \includegraphics[width=1\textwidth,, ]{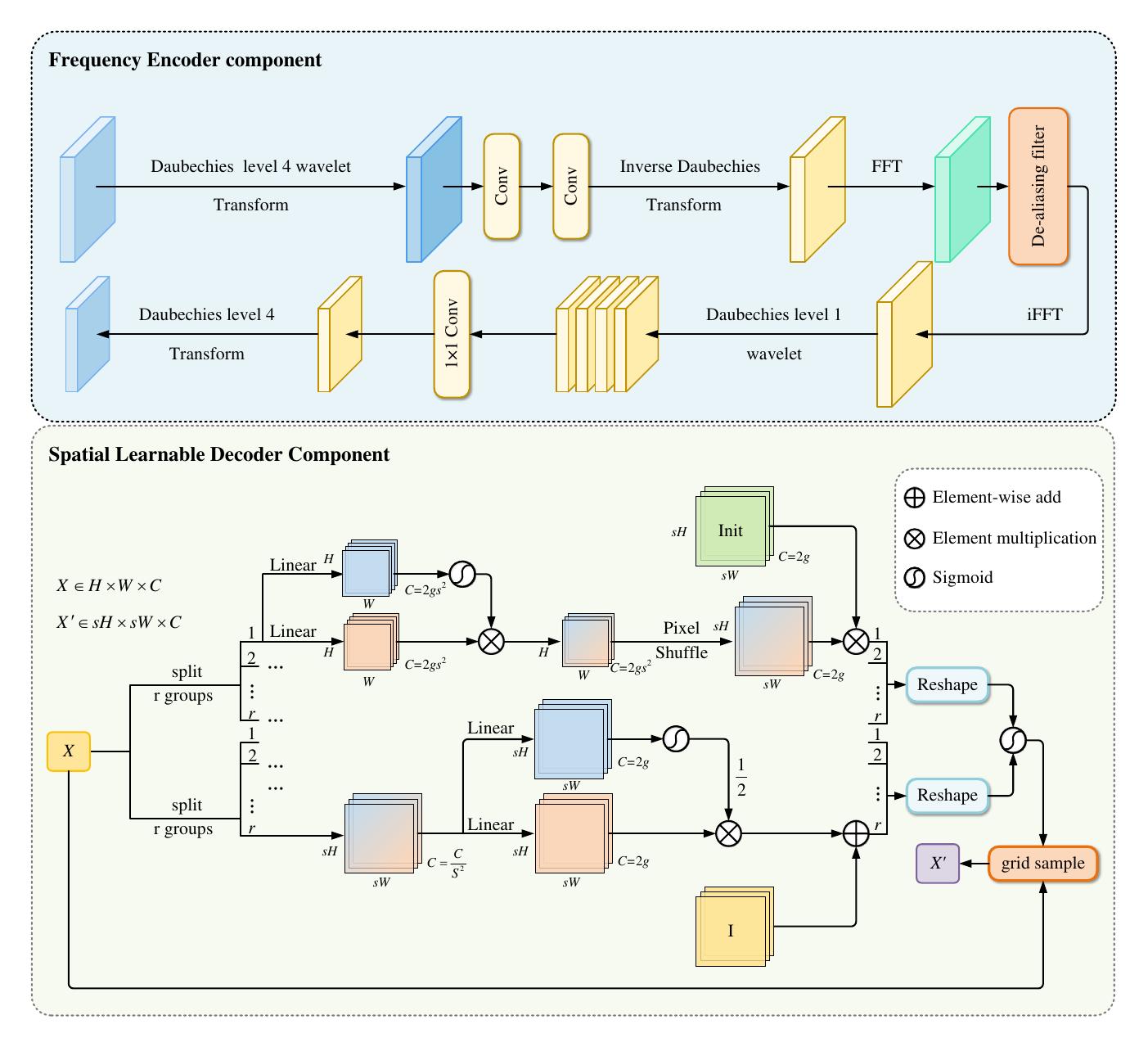} 
    \caption{The Structure of Component of Frequency domain-based Encoder and Spatial Learnable Decoder. The Encoder component shows both the convolution and downsampling process. Structure of Decoder mainly shows the upsampling component. }
    \label{fig:Main progress}
\end{figure*}

\section{Related Work} \label{sec:rw}

\subsection{Medical Image Segmentation}

Medical image segmentation is a foundational task in computer-aided diagnosis and surgical planning. Classical architectures like U-Net~\cite{ronneberger2015u} and its numerous variants~\cite{li2018h,8932614,huang2020unet} have shown remarkable success in pixel-level prediction by leveraging multi-scale skip connections and symmetric encoder-decoder designs. Advanced extensions such as Residual U-Net~\cite{zhang2018road} and nnU-Net~\cite{isensee2021nnu} further improve performance through optimized residual pathways and automatic configuration tuning, achieving strong generalization across various medical imaging modalities.
However, most CNN-based architectures rely heavily on repeated convolution and pooling operations, which inevitably cause spatial resolution loss and aliasing artifacts. This hampers the accurate segmentation of fine structures, especially for small or irregular anatomical regions. Additionally, conventional skip connections that directly sum encoder and decoder features may result in limited fusion of hierarchical semantics, hindering segmentation precision.
Transformer-based segmentation models have recently gained momentum due to their ability to capture long-range dependencies. Architectures such as UNETR~\cite{hatamizadeh2022unetr} and Swin-UNet~\cite{cao2022swin} utilize self-attention mechanisms to model global context effectively. Hybrid models like TransUNet~\cite{chen2021transunet} and nnFormer~\cite{zhou2021nnformer} combine CNN's inductive bias with Transformer’s capacity for contextual modeling. Nevertheless, these methods often depend on large-scale pretraining (e.g., ImageNet), which may not align with the distribution of medical imaging data and introduces significant computational overhead—limiting their practical deployment in clinical environments.

\subsection{Frequency-Based Encoder}

Incorporating frequency-domain information has emerged as a promising direction for enhancing feature extraction in segmentation models. Wavelet-based techniques have been widely adopted for downsampling, receptive field expansion, and feature compression. For instance, two-level wavelet transforms are employed to compress high-resolution features and enlarge receptive fields~\cite{finder2024wavelet,finder2022wavelet}, while Haar wavelet transforms are used in semantic segmentation downsampling~\cite{xu2023haar} and inverse wavelet transforms in upsampling~\cite{liu2018multi}. Additionally, wavelet pooling has been proposed to replace traditional spatial pooling~\cite{duan2017sar,li2018h}.
Despite their benefits, most of these methods treat wavelet and frequency components as auxiliary modules, without integrating them into the core network design. Recent studies in image generation have shown that frequency-conditioned architectures can enhance fine-grained structural control~\cite{shen2025imaggarment}. Inspired by this, we aim to unify wavelet and frequency-aware mechanisms directly into the encoder path, facilitating both spectral compression and discriminative representation learning in high-resolution medical images.

\subsection{Spatial Learnable Decoder and Loss Function}

Recent research has explored lightweight and content-aware upsampling modules to mitigate checkerboard artifacts and preserve object boundaries. Dynamic upsampling operators such as FADE~\cite{lu2022fade} and CARAFE~\cite{wang2019carafe} introduce learnable fusion strategies for decoder-encoder integration. Moreover, adaptive sampling~\cite{liu2023learning} and frequency-aware filtering~\cite{chen2024frequency} have been employed to enhance structural fidelity in reconstructed maps. These approaches have demonstrated improvements in fine detail recovery, particularly in complex medical scenes.
Loss functions also play a crucial role in guiding segmentation performance. Generalized Dice Loss~\cite{sudre2017generalised} addresses class imbalance by weighting contributions from minority regions, while Top-K Cross-Entropy Loss~\cite{fan2017learning} encourages learning from difficult pixels. In more recent work, controllable generation frameworks have employed specialized decoders to enhance spatial expressiveness in high-resolution generation tasks~\cite{shen2025imagdressing}, while pose-conditional generation models like IMAGPose~\cite{shen2024imagpose} and progressive diffusion-based approaches~\cite{shen2023advancing} further highlight the importance of structure-aware reconstruction. Moreover, motion-aware temporal diffusion models~\cite{shen2025long} have demonstrated that injecting prior knowledge into decoder stages significantly improves temporal and spatial coherence. These insights motivate our design of a spatial learnable decoder combined with frequency-guided supervision to improve segmentation precision under complex imaging conditions.

\section{Proposed Method}\label{sec:method} 
Motivated by the challenges of spatial-frequency aliasing and imbalance classes segmentation performance in existing medical segmentation models, we propose the Frequency domain-based FreqU-FNet innovatively integrates frequency-domain anti-aliasing, hierarchical adaptive upsampling and refined feature fusion methods into a unified CNN-based segmentation architecture. Fig.~\ref{fig:Main progress} provides an illustrative overview of our architecture.

\subsection{Overview}
Given an input medical image, our segmentation network extracts features through an encoder-decoder paradigm enhanced by frequency-domain processing and sophisticated multi-scale feature integration. The architecture consists of three main components: a frequency domain-based encoder with anti-aliasing capabilities,  an spatial learnable decoder (SLD) with adaptive hierarchical multi-branch upsampling, and a spatial auxiliary learning module with a modified loss function tailored for frequency domain learning.
\subsection{Frequency Domain Encoder}  

Traditional CNN pooling operations introduce spatial-frequency aliasing, causing the loss of high-resolution details essential for medical image segmentation. Our encoder addresses this by combining frequency domain low-pass filtering convolution (FLC)  and Daubechies wavelet downsampling.

Given an input feature map X,we first apply the Daubechies (db) wavelet transform for multi-resolution decomposition.
\begin{equation}
    \{X_{LL}, X_{LH}, X_{HL}, X_{HH}\} = \text{WT}_{db}(X).
\end{equation}
Here, $X$ denotes the input feature map. $\mathrm{WT}_{db}(\cdot)$ is the Daubechies wavelet transform operator. $X_{LL}$ represents the low-frequency approximation coefficients, while $X_{LH}$, $X_{HL}$, and $X_{HH}$ denote the horizontal, vertical and diagonal high-frequency detail coefficients respectively.
\begin{equation}
    X_{reconstructed} = \text{IWT}_{db}(X_{LL}, X_{LH}, X_{HL}, X_{HH}).
\end{equation}
Where $X_{\mathrm{reconstructed}}$ is the feature map reconstructed by the inverse wavelet transform, and $\mathrm{IWT}_{db}(\cdot)$ stands for the inverse Daubechies wavelet transform operator.

For more precise frequency aliasing elimination in the Fourier domain, we first perform the wavelet inverse transform on specific subbands, followed by the Fourier transform:
\begin{equation}
    X_{freq}=\mathcal F(X).
\end{equation}
$\mathcal{F}(\cdot)$ denotes the two-dimensional Fourier transform, and $X_{\mathrm{freq}}$ is the representation of the feature map in the frequency domain.
In the Fourier domain, we design and apply a low-pass filter $M$ to selectively remove high-frequency noise and aliasing artifacts:We then apply a low-pass filter in the frequency domain using a mask $M$: $M$.
\begin{equation}
M(u,v) = 
\begin{cases} 
1, & \text{if} \quad |u - u_c| \leq \tau h,\quad |v - v_c| \leq \tau w \\
0, & \text{otherwise}
\end{cases} .
\end{equation}
where $M(u, v)$ is the value of the low-pass filter mask at frequency coordinates $(u, v)$. $(u_c, v_c)$ represent the center coordinates of the frequency map. $h$ and $w$ are the height and width of the frequency map, and $\tau$ is a ratio parameter ($0 < \tau < 1$) that controls the size of the preserved low-frequency region. The filtered features are transformed back to the spatial domain by:
\begin{equation}
    X_{low} = \mathcal{F}^{-1}(X_{freq} \odot M).
\end{equation}
Here $\mathcal{F}^{-1}(\cdot)$ is the inverse two-dimensional Fourier transform. $X_{\mathrm{freq}}$ is the frequency domain feature map, $M$ is the low-pass filter mask, $\odot$ denotes element-wise multiplication, and $X_{\mathrm{low}}$ is the filtered feature map transformed back into the spatial domain.

Finally, we reapply the Daubechies wavelet transform for downsampling, preserving the key information after frequency domain processing:

\begin{equation}
    \{X_{LL}^{down}, X_{LH}^{down}, X_{HL}^{down}, X_{HH}^{down}\} = \text{WT}_{db}(X_{low}).
\end{equation}
Here \(X_{LL}^{\mathrm{down}}, X_{LH}^{\mathrm{down}}, X_{HL}^{\mathrm{down}}, X_{HH}^{\mathrm{down}}\) are the approximation and detail subbands obtained by applying the Daubechies wavelet transform \(\mathrm{WT}_{db}\) a second time to the spatial-domain feature map \(X_{\mathrm{low}}\), effectively downsampling while preserving critical frequency information.
Overall, this hybrid approach can be formalized as a mapping function $\Phi$:
\begin{equation}
    \Phi (X) = \text{WT}_{db}(\mathcal{F}^{-1}(\mathcal F (\text{IWT}_{db}(\text{WT}_{db}(X)))\odot M).
\end{equation}
In this composite mapping, \(\Phi(\cdot)\) denotes the entire frequency-domain encoder function combining wavelet decomposition (\(\mathrm{WT}_{db}\)), inverse wavelet reconstruction (\(\mathrm{IWT}_{db}\)), Fourier transform (\(\mathcal{F}\)), low-pass filtering by \(M\), inverse Fourier transform (\(\mathcal{F}^{-1}\)), and a final wavelet transform.
Through this mapping, we combine the multi-scale analysis capabilities of wavelet transform with the global frequency filtering advantages of Fourier transform, effectively reducing frequency aliasing artifacts while preserving critical anatomical details in specific frequency bands, especially for minority classes primarily existing within specific frequency ranges. Experiments show that this method performs excellently in handling imbalanced class problems in medical image segmentation, capturing the signal features of minority classes better than traditional spatial domain convolutional neural networks.

\subsection{Spatial learnable decoder (SLD)}  
Traditional CNN-based decoder architectures employ straightforward bilinear interpolation or deconvolution for upsampling, leading to whether loss of critical details or checkboard artifact ~\cite{aitken2017checkerboard}. Our SLD tackles these issues by incorporating a dedicated \textit{adaptive hierarchical multi-branch upsampling} that provides sophisticated multi-scale context aggregation and detail preservation through dynamically combined upsampling pathways. \\

The adaptive upsampling component inside SLD employs two complementary upsampling pathways that are adaptively fused based on learned importance weights. Given an input feature map $\mathbf{X} \in \mathbb{R}^{B \times C \times H \times W}$, where $B$ is the batch size, $C$ is the number of channels, and $H \times W$ are the spatial dimensions, our module generates an upsampled output $\mathbf{Y} \in \mathbb{R}^{B \times C\times sH \times sW}$, where $s$ is the scaling factor and $C_{out}$ is the desired output channel dimension.

\subsubsection{Native-Space Dynamic Sampling pathway.} 
Native-Space Dynamic Sampling Sampling pathway aim to learn fine-grained local deformation directly form the original resolution feature. First, the input $X\in R^{B\times C \times H \times W}$ is split into $r$ groups along the channel dimension, each group calculate a base coordinate grid $I\in R^{B\times 2g\times sH \times sW}$ to initialize sampling positions. Next, $X$ was sent through two parallel linear layer, one layer pathway is to adjust channels to $R^{B \times 2gs^2 \times B\times W}$, where $s$ is the sampling point numbers. Another one will used to modulate offset. The two pathways then multiply and reshaped by pixel shuffle to $2\times H \times W$. The offset was then added onto the original sample grid, formulating the sampling set.The process can be briefly expressed in the formula below:
\begin{equation}
     O = I + \frac{1}{2}sigmoid(linear_1(X))\cdot linear_2(X).
\end{equation}
Here, \(O\) is the learned coordinate offset for grid sampling; \(I\) is the base sampling grid; \(\mathrm{linear}_1\) and \(\mathrm{linear}_2\) are two parallel fully-connected layers applied to the decoder input feature map \(X\); \(\mathrm{sigmoid}(\cdot)\) normalizes offsets into \([0,1]\); and the element-wise multiplication \(\cdot\) modulates the offsets.
The coordinates $\mathbf{O}_{NSD}$ are then used for grid sampling from the input features and rearranged to produce the upsampled output $F_1$.

\subsubsection{Space-Channel Exchange Sampling pathway.}
Space-Channel Exchange Sampling pathway learning Employs channel rearrangement for efficient upsampling. We first split feature into $r$ groups.Unlike $F_1$, here we do the pixel shuffle first, reshaping the input to $\frac{C}{s^2}$,the channel are rearranged to reduce dimensionality and upsample spatially.Then the input was upsampled to $2g \times sH\times sW$, in same equation, we modulate offset and generate $F_2$.
After that, we perform deformable sampling guided by $O$, concatenate all the groups together.The final output is produced by dynamically weighting and combining the outputs from both pathways.

Lastly, a $1\times1$ convolutional layer adjusts the channel dimensions to match the desired output. The dual complementary pathways capture different aspects of features and adaptive fusion optimizes information flow for each spatial location.

For each stage of Spatial Learnable Decoder, the decoder first expand the features by the proposed adaptive hierarchical multi-branch upsampling, then the expanded feature map are concatenated with the corresponding feature maps from the encoder at the same resolution. This supplies high-resolution detail to further prevent any detail loss in downsampling. After concentation, two $3\times3$ convolutions are applied. Each convolution is followed by instance normalization and LeakyReLU activation function. Instance normalization normalizes per-sample and helps when batch sizes vary and LeakyReLU keeps a small gradient for negative inputs and stablizes training.

\subsection{Loss Function}

To better capture fine anatomical boundaries and address class imbalance in medical image segmentation, we propose a composite loss function incorporating a frequency-aware loss (FAL), a multi-class Dice loss, and an average Top-K cross-entropy loss.

\noindent\textbf{Frequency-Aware Loss.} To emphasize high-frequency structures such as lesion edges, we introduce a wavelet-domain loss computed via Discrete Wavelet Transform (DWT) using Daubechies basis. Both the predicted probability map $\hat{y}$ and the ground-truth mask $y$ are decomposed into low-frequency ($L$) and high-frequency ($H$) components:
\begin{equation}
\mathcal{W}(\hat{y}) = (\hat{y}_L, \hat{y}_H), \quad \mathcal{W}(y) = (y_L, y_H),
\end{equation}
\noindent where $\mathcal{W}(\cdot)$ denotes DWT. The frequency loss is then defined as the L1 norm over high-frequency subbands:
\begin{equation}
\mathcal{L}_{\text{Freq}} = \frac{1}{D_H} \sum_{d=1}^{D_H} \|\hat{Y}_{H}^{d} - Y_{H}^{d}\|_1,
\end{equation}
\noindent where $D_H$ is the number of subbands (e.g., $3$ for 2D: LH, HL, HH) and the norm is computed over all spatial dimensions and channels. We omit the LL (low-low) component as it mainly encodes coarse structural information already captured by conventional losses.

\noindent\textbf{Dice Loss.} We adopt the multi-class Dice loss~\cite{drozdzal2016importance}, defined as:
\begin{equation}
\mathcal{L}_{\text{Dice}} = - \frac{2}{|K|} \sum_{k \in K} \frac{\sum_{i \in I} u_i^k v_i^k}{\sum_{i \in I} u_i^k + \sum_{i \in I} v_i^k},
\end{equation}
\noindent where $u_i^k$ and $v_i^k$ are the predicted and ground-truth class probabilities for pixel $i$ and class $k$.

\noindent\textbf{Top-K Cross-Entropy Loss.} To emphasize hard examples, we incorporate the average Top-K cross-entropy loss~\cite{fan2017learning}. First, per-pixel cross-entropy is computed as:
\begin{equation}
\mathcal{L}_{CE}(x_i, y_i) = -\frac{1}{C} \sum_{j=1}^{C} \log \left( \frac{\exp(x_{i,j})}{\sum_{k=1}^{C} \exp(x_{i,k})} \right),
\end{equation}
\noindent where $x_{i,j}$ is the logit for class $j$ and $y_i$ is the ground-truth label. The Top-K loss is defined as the average over the hardest $k\%$ samples:
\begin{equation}
\mathcal{L}_{\text{TopK}} = \frac{1}{K} \sum_{i \in \mathcal{I}_{\text{TopK}}} L_i, \quad K = \left\lfloor N \cdot \frac{k}{100} \right\rfloor,
\end{equation}
\noindent where $\mathcal{I}_{\text{TopK}}$ denotes the index set of the top $K$ losses.

\noindent\textbf{Final Loss.} The total loss is a weighted combination of the three components:
\begin{equation}
\mathcal{L}_{\text{Total}} = w_{\text{Dice}} \cdot \mathcal{L}_{\text{Dice}} + w_{\text{TopK}} \cdot \mathcal{L}_{\text{TopK}} + w_{\text{Freq}} \cdot \mathcal{L}_{\text{Freq}},
\end{equation}
\noindent where $w_{\text{Dice}}, w_{\text{TopK}}, w_{\text{Freq}}$ are hyperparameters. This combination enables the model to simultaneously focus on class balance, boundary precision, and difficult regions, ultimately improving segmentation robustness and accuracy.

\section{Experiment and Analysis}\label{sec:exp}  
Extensive experiments demonstrate that FreqU-FNet consistently outperforms existing methods across three large-scale datasets—MSD-Prostate, MSD-Pancreas, and MSD-Lung—highlighting its superior generalization and segmentation accuracy.

\subsection{Datasets}
\textbf{\emph{MSD Prostate Dataset}}
The prostate is a part of the male reproductive system, located below the bladder and surrounded by the initial urethra. Its main function is to produce prostatic fluid. Multimodal magnetic resonance imaging (MRI) helps to assess the anatomy of the prostate, including the central gland and peripheral zone. These two regions have different anatomical and physiological characteristics in the prostate. The central gland mainly contains prostatic ducts, while the peripheral zone is the most common site of prostate cancer. By segmenting these two regions, doctors can more precisely detect and assess prostate diseases, such as prostate cancer, and other problems that may affect prostate health, enabling early diagnosis and treatment.The goal of this dataset is to segment the central gland and thee peripheral zone from multimodel MR(T2, ADC) images. We chose this dataset because this dataset need to segment two adjacent regions, which have great variability among different individuals.\\
\textbf{\emph{MSD Pancreas Dataset}}
Pancreatic cancer is a lethal malignant tumor. Its eraly symptoms are often not obvious, resulting in having entered an advanced stage at the time of diagnosis. The survival rate for pancreatic cancer is poor, with five-year survival rates usually less than 10\%. Early detection and diagnosis are essential to improve the survival rate of patients with pancreatic cancer. Imaging examinations such as CT and MRI play a central role in the early diagnosis and staging of pancreatic cancer. Through imaging segmentation, doctors can better understand the location,, size and metastasis of the tumor, so as to make appropriate treatment plan.The dataset contains three types of pancreatic tumors. We chose this dataset because the imbalance label, which includes large background, medium pancreas and small tumor structures. \\
\textbf{\emph{MSD Lung Dataset}}
Lung cancer ranks among the leading causes of cancer mortality, and contrast-enhanced CT imaging provides detailed 3D views of pulmonary anatomy and tumor masses. The MSD Lung dataset comprises chest CT scans with voxel-wise annotations of non-small cell lung tumors, captured under a variety of scanner models and imaging protocols. We selected this dataset because lung tumors often present as small, irregular lesions within a vast background of healthy tissue, creating an extreme class imbalance that rigorously tests a model’s sensitivity to subtle pathology and precision at complex boundaries .

\subsection{Evaluation Metrics}  
To quantitatively assess segmentation quality, we adopt two complementary metrics: the Dice similarity coefficient (DSC) and the class‐wise Dice difference (or “Gap”). These metrics are widely used in medical image segmentation and together provide both an absolute measure of overlap and an indication of class imbalance robustness.
The DICE measures the overlap between the predicted mask $P$ and the ground-truth mask $G$. For a single class $k$, it is defined as:

\begin{equation}
    \mathrm{DICE}_k \;=\; \frac{2\,|P_k \,\cap\, G_k|}{|P_k| + |G_k|}\;\in[0,1].
\end{equation}
where $\mathrm{DICE}_k$ denotes the DICE score of class $k$, $|\cdot|$ denotes the cardinality (i.e.\ number of pixels or voxels). A DICE of 1 indicates perfect agreement, while 0 indicates no overlap. We choose DICE because it directly reflects the spatial overlap critical in clinical applications, being insensitive to true‐negative abundance in large background regions and is the standard in recent medical segmentation challenges, facilitating fair comparison.\\
To evaluate how evenly a method performs across majority and minority classes, we define the Dice difference as the range between the highest and lowest class DICE.A smaller Gap indicates more balanced performance, presenting a model’s ability to handle underrepresented (often small or low‐contrast) structures without overfitting to large regions. Since one of our core contributions is to reduce frequency‐aliasing and boost minority‐class accuracy, reporting the Dice difference directly demonstrates how FreqU-FNet narrows the performance disparity that standard spatial‐domain networks exhibit. Together, DICE and Gap comprehensively support our work by measuring not only overall accuracy but also the consistency of class‐level segmentation.
\subsection{Implementation Details} 
We first crop all the data to the region of nonzero value as it has no effect on these dataset while reducing the size the data. The data of patients are then resampled to the median voxel spacing of the respective dataset, where the image data were interpolated using third-order spline and the corresponding segmentation masks were interpolated using nearest neighbor. All data are normalized according to the mean and standard deviation of corresponding datasets. For data augmentation, random rotations, random scaling, random elastic deformations, gamma correction augmentation and mirroring techniques were applied during training. Models are trained and evaluated using five-fol cross-validation on the training set. We use Adam optimizer with an initial learning rate of $4\times10^{-3}$. We keep an expoential moving average of the validation and training losses. The learning rate was reduced linearly by factor 2 and the minimum learning rate was set to $10^{-6}$.

\subsection{Comparison with State-of-the-art Methods} 

\subsubsection{Comparsion on MSD Prostate Dataset}

\begin{table}[t]
\centering
\caption{Comparsion on MSD Prostate Dataset}
\scalebox{0.95}{
\scriptsize
\begin{tabular}{l|cc|c}
\hline
Method & DICE 1 & DICE 2 & DICE Diff \\
\hline
2D nnU-Net ~\cite{isensee2021nnu} & 61.98 & 84.31 & 22.33 \\
3D nnU-Net ~\cite{isensee2021nnu} & 60.77 & 83.73 & 22.96 \\
Ensemble 2D nnUNet + 3D nnUNet ~\cite{isensee2021nnu} & 63.78 & 85.31 & 21.53 \\
AttnUNet ~\cite{cheng2025effective} & 72.77 & 86.53 & \textbf{13.76} \\
Swin-UNet ~\cite{cao2022swin} & 75.11 & 88.93 & 14.96 \\
UNETR ~\cite{hatamizadeh2022unetr}& 73.37 & 88.93 & 15.56 \\
TransUNet ~\cite{chen2021transunet} & \textbf{73.42} & 87.93 & 14.35 \\
\hline
FreqU-FNet & 73.21 & \textbf{90.08} & 18.49 \\
\hline
\end{tabular}
}
\label{tab:prostate_results}
\end{table}
Table 1 shows that FreqU-FNet achieves a balanced improvement on both gland classes. Specifically, our method attains a 90.08 \% Dice on the peripheral zone, 2.15 points higher than TransUNet’s 87.93 \% \cite{chen2021transunet} and 3.55 points above Swin-UNet’s 86.53 \% \cite{cao2022swin}, while maintaining central gland accuracy at 73.21 \%, on par with the best hybrid CNN-Transformer approaches. This result indicates that the frequency, domain encoder’s combined wavelet–Fourier filtering effectively suppresses aliasing and sharpens low-contrast boundaries, enabling the network to delineate the peripheral zone more faithfully than purely spatial or self-attention models.

\subsubsection{Comparsion on MSD Pancreas Dataset}

\begin{table}[t]
\centering

\caption{Comparsion on MSD Pancreas Dataset}
\scalebox{0.95}{
\scriptsize
\begin{tabular}{l|cc|c}
\hline
Method & DICE (1) & DICE (2) & DICE Diff \\
\hline
2D nnU-Net ~\cite{isensee2021nnu} & 74.70  & 35.41  & 39.29 \\
3D nnU-Net ~\cite{isensee2021nnu} & 77.69 & 42.69 & 35.00 \\
Ensemble 2D nnUNet + 3D nnUNet ~\cite{isensee2021nnu} & 79.30 & 52.12 & 27.18 \\
AttnUNet ~\cite{cheng2025effective} & 78.65 & 51.15 & 27.50 \\
Swin-UNet ~\cite{cao2022swin} & 80.26 & 54.51 & 25.75 \\
UNETR ~\cite{hatamizadeh2022unetr}& \textbf{81.55} & 53.32 & 28.23 \\
TransUNet  ~\cite{chen2021transunet} & 80.66 & 54.11 & 26.55 \\
\hline
FreqU-FNet & 79.02 & \textbf{63.38} & \textbf{15.64} \\
\hline
\end{tabular}
}
\label{tab:pancreas_results}
\end{table}
Table~\ref{tab:pancreas_results} demonstrates that FreqU-FNet effectively mitigates class imbalance on the MSD Pancreas dataset. It achieves a tumor Dice of 63.38\%, outperforming Swin-UNet (54.51\%)~\cite{cao2022swin} and TransUNet (54.11\%)~\cite{chen2021transunet} by over 8.8 points, while maintaining a competitive 79.02\% Dice on healthy pancreas. The Dice gap is reduced to 15.64\%, nearly halving the 27\%--39\% range observed in nnU-Net variants~\cite{isensee2021nnu}.
This improvement stems from our Spatial Learnable Decoder (SLD), which combines a native-space path for sub-pixel deformation modeling and a space-channel path for global structure preservation, guided by learned fusion weights. Additionally, the frequency-combined loss emphasizes high-frequency wavelet discrepancies, encouraging finer tumor boundary refinement beyond standard Dice or cross-entropy terms.

\subsubsection{Comparisons on MSD Lung Dataset}

\begin{table}[t]
\centering
\caption{Comparsion on MSD Lung Dataset}
{
\scriptsize
\begin{tabular}{l|cc|c}
\hline
Method & DICE (1) \\
\hline
2D nnU-Net ~\cite{isensee2021nnu} & 74.70     \\
3D nnU-Net ~\cite{isensee2021nnu} & 77.69   \\
Ensemble 2D nnUNet + 3D nnUNet ~\cite{isensee2021nnu} & 79.30   \\
AttnUNet ~\cite{cheng2025effective} & 78.65   \\
Swin-UNet ~\cite{cao2022swin}  & 80.26   \\
UNETR ~\cite{hatamizadeh2022unetr} & 73.54  \\
TransUNet  ~\cite{chen2021transunet} & 77.00   \\
\hline
FreqU-FNet & \textbf{80.77}   \\
\hline
\end{tabular}
}
\label{tab:lung_results}
\end{table}
Table 3 (MSD Lung dataset) demonstrates that FreqU-FNet achieves the highest overall lung Dice of 80.77 \%, surpassing Swin-UNet’s 80.26 \% \cite{cao2022swin} and TransUNet’s 77.00 \% \cite{chen2021transunet}. Beyond gross mask accuracy, our model excels at tracing thin airway branches and lesion margins that conventional upsampling or self-attention often blur. The frequency-domain encoder filters out CT speckle noise while reinforcing vessel and nodule signals and the decoder’s deformable sampling avoids checkerboard artifacts—together producing crisp, contiguous delineations even in challenging scan conditions.

\subsection{Ablation Studies and Analysis} 

\begin{table}[t]
  \centering\small
  \caption{Ablation Study on the MSD Pancreas Dataset}
  \label{tab:ablation_processed}
  \begin{tabular}{lccccccc}
    \toprule
    Variant       & FLC & DB  Downsampling & SLD & FAL & DICE\,1 (\%) & DICE\,2 (\%) & Gap (\%) \\
    \midrule
     FLC       & \xmark          & \cmark      & \cmark      & \cmark      &  76.88       &     60.14     &    16.74      \\
     DB Downsampling                     & \cmark          & \xmark  &\ \cmark          & \cmark      & 76.05        & 62.57        & 13.48         \\
     SLD                     & \xmark          & \cmark     & \cmark       & \cmark      &     78.50     & 57.76        &   20.74     \\
     FAL      & \cmark          & \cmark      & \cmark      & \xmark      & 74.93        & 60.11        &     14.82    \\
     \hline
     FreqU-FNet & \cmark &    \cmark   &    \cmark   &    \cmark  &  79.02   &  63.38   & 15.64 \\

    \bottomrule
  \end{tabular}
\end{table}
To quantify the contribution of each component in our FreqU-FNet framework, we conducted an ablation study on the MSD Pancreas dataset, evaluating segmentation performance with two DICE scores (DICE 1, DICE 2) and their difference (Gap) under identical training settings (Table \ref{tab:ablation_processed}).

Firstly, removing the Frequency-Level Calibration (FLC) module causes the largest drop in DICE 1 (-2.14 \%) and DICE 2 (–3.24 \%) relative to the full model, and increases the DICE gap by 1.10 \%, demonstrating that FLC is essential for harmonizing multi-scale frequency information. Secondly, omitting the DB downsampling block yields a modest decrease in DICE 1 (-2.97 \%) but only -0.81 \% in DICE 2, and actually narrows the gap to 13.48 \%, indicating that downsampling chiefly benefits coarse structure capture while slightly improving class balance. Thirdly, without the Spatial–Local Detail (SLD) module, DICE 2 plunges by 5.62 \% and the gap widens by 5.10 \%, confirming that SLD is critical for fine boundary delineation at the cost of a more pronounced inter-class disparity. Finally, excluding the Frequency Attention Layer (FAL) hurts both DICE 1 (-4.09 \%) and DICE 2 (-3.27 \%) and reduces the gap modestly, showing FAL’s role in balancing global versus local features. Overall, the intact FreqU-FNet—which integrates FLC, DB downsampling, SLD and FAL—achieves the best trade-off, with 79.02 \% DICE 1, 63.38 \% DICE 2 and a gap of 15.64 \%, outperforming each ablated variant and confirming that every component meaningfully contributes to segmentation accuracy and stability.

\section{Conclusion}\label{sec:con} 
In this paper, we introduced FreqU-FNet, a novel frequency domain-based segmentation architecture that effectively addresses the challenges of medical image segmentation. Our approach combines Frequency Low-pass Convolution with Daubechies wavelet downsampling to reduce frequency aliasing artifacts while enhancing feature extraction for minority classes. The Spatial Learnable Decoder combines frequency domain information and efficiently recovers spatial information through adaptive weighting of distinct upsampling pathways. Our frequency-enhanced loss function guides the model to learn effectively in the frequency domain while addressing class imbalance problems. Experimental results demonstrate that FreqU-FNet consistently outperforms state-of-the-art CNN and Transformer-based models in segmentation accuracy, particularly in reducing performance gaps between majority and minority classes.

%
%
%
%

\bibliographystyle{splncs04}

\bibliography{paper}

\begin{thebibliography}{10}
\providecommand{\url}[1]{\texttt{#1}}
\providecommand{\urlprefix}{URL }
\providecommand{\doi}[1]{https://doi.org/#1}

\bibitem{aitken2017checkerboard}
Aitken, A., Ledig, C., Theis, L., Caballero, J., Wang, Z., Shi, W.: Checkerboard artifact free sub-pixel convolution: A note on sub-pixel convolution, resize convolution and convolution resize. arXiv preprint arXiv:1707.02937  (2017)

\bibitem{cao2022swin}
Cao, H., Wang, Y., Chen, J., Jiang, D., Zhang, X., Tian, Q., Wang, M.: Swin-unet: Unet-like pure transformer for medical image segmentation. In: European conference on computer vision. pp. 205--218. Springer (2022)

\bibitem{chen2021transunet}
Chen, J., Lu, Y., Yu, Q., Luo, X., Adeli, E., Wang, Y., Lu, L., Yuille, A.L., Zhou, Y.: Transunet: Transformers make strong encoders for medical image segmentation. arXiv preprint arXiv:2102.04306  (2021)

\bibitem{chen2024frequency}
Chen, L., Fu, Y., Gu, L., Yan, C., Harada, T., Huang, G.: Frequency-aware feature fusion for dense image prediction. IEEE Transactions on Pattern Analysis and Machine Intelligence  (2024)

\bibitem{chen2024semantic}
Chen, L., Gu, L., Fu, Y.: When semantic segmentation meets frequency aliasing. arXiv preprint arXiv:2403.09065  (2024)

\bibitem{cheng2025effective}
Cheng, Z., Yuan, D., Zhang, W., Lukasiewicz, T.: Effective and efficient medical image segmentation with hierarchical context interaction. In: 2025 IEEE/CVF Winter Conference on Applications of Computer Vision (WACV). pp. 9396--9405. IEEE (2025)

\bibitem{drozdzal2016importance}
Drozdzal, M., Vorontsov, E., Chartrand, G., Kadoury, S., Pal, C.: The importance of skip connections in biomedical image segmentation. In: International Workshop on Deep Learning in Medical Image Analysis. pp. 179--187. Springer (2016)

\bibitem{duan2017sar}
Duan, Y., Liu, F., Jiao, L., Zhao, P., Zhang, L.: Sar image segmentation based on convolutional-wavelet neural network and markov random field. Pattern Recognition  \textbf{64},  255--267 (2017)

\bibitem{fan2017learning}
Fan, Y., Lyu, S., Ying, Y., Hu, B.: Learning with average top-k loss. Advances in neural information processing systems  \textbf{30} (2017)

\bibitem{finder2024wavelet}
Finder, S.E., Amoyal, R., Treister, E., Freifeld, O.: Wavelet convolutions for large receptive fields. In: European Conference on Computer Vision. pp. 363--380. Springer (2024)

\bibitem{finder2022wavelet}
Finder, S.E., Zohav, Y., Ashkenazi, M., Treister, E.: Wavelet feature maps compression for image-to-image cnns. Advances in Neural Information Processing Systems  \textbf{35},  20592--20606 (2022)

\bibitem{hatamizadeh2022unetr}
Hatamizadeh, A., Tang, Y., Nath, V., Yang, D., Myronenko, A., Landman, B., Roth, H.R., Xu, D.: Unetr: Transformers for 3d medical image segmentation. In: Proceedings of the IEEE/CVF winter conference on applications of computer vision. pp. 574--584 (2022)

\bibitem{huang2020unet}
Huang, H., Lin, L., Tong, R., Hu, H., Zhang, Q., Iwamoto, Y., Han, X., Chen, Y.W., Wu, J.: Unet 3+: A full-scale connected unet for medical image segmentation. In: ICASSP 2020-2020 IEEE international conference on acoustics, speech and signal processing (ICASSP). pp. 1055--1059. IEEE (2020)

\bibitem{isensee2021nnu}
Isensee, F., Jaeger, P.F., Kohl, S.A., Petersen, J., Maier-Hein, K.H.: nnu-net: a self-configuring method for deep learning-based biomedical image segmentation. Nature methods  \textbf{18}(2),  203--211 (2021)

\bibitem{isensee2024nnu}
Isensee, F., Wald, T., Ulrich, C., Baumgartner, M., Roy, S., Maier-Hein, K., Jaeger, P.F.: nnu-net revisited: A call for rigorous validation in 3d medical image segmentation. In: International Conference on Medical Image Computing and Computer-Assisted Intervention. pp. 488--498. Springer (2024)

\bibitem{li2018h}
Li, X., Chen, H., Qi, X., Dou, Q., Fu, C.W., Heng, P.A.: H-denseunet: hybrid densely connected unet for liver and tumor segmentation from ct volumes. IEEE transactions on medical imaging  \textbf{37}(12),  2663--2674 (2018)

\bibitem{liu2018multi}
Liu, P., Zhang, H., Zhang, K., Lin, L., Zuo, W.: Multi-level wavelet-cnn for image restoration. In: Proceedings of the IEEE conference on computer vision and pattern recognition workshops. pp. 773--782 (2018)

\bibitem{liu2023learning}
Liu, W., Lu, H., Fu, H., Cao, Z.: Learning to upsample by learning to sample. In: Proceedings of the IEEE/CVF International Conference on Computer Vision. pp. 6027--6037 (2023)

\bibitem{lu2022fade}
Lu, H., Liu, W., Fu, H., Cao, Z.: Fade: Fusing the assets of decoder and encoder for task-agnostic upsampling. In: European Conference on Computer Vision. pp. 231--247. Springer (2022)

\bibitem{ronneberger2015u}
Ronneberger, O., Fischer, P., Brox, T.: U-net: Convolutional networks for biomedical image segmentation. In: Medical image computing and computer-assisted intervention--MICCAI 2015: 18th international conference, Munich, Germany, October 5-9, 2015, proceedings, part III 18. pp. 234--241. Springer (2015)

\bibitem{shen2025imagdressing}
Shen, F., Jiang, X., He, X., Ye, H., Wang, C., Du, X., Li, Z., Tang, J.: Imagdressing-v1: Customizable virtual dressing. In: Proceedings of the AAAI Conference on Artificial Intelligence. vol.~39, pp. 6795--6804 (2025)

\bibitem{shen2024imagpose}
Shen, F., Tang, J.: Imagpose: A unified conditional framework for pose-guided person generation. Advances in neural information processing systems  \textbf{37},  6246--6266 (2024)

\bibitem{shen2025long}
Shen, F., Wang, C., Gao, J., Guo, Q., Dang, J., Tang, J., Chua, T.S.: Long-term talkingface generation via motion-prior conditional diffusion model. arXiv preprint arXiv:2502.09533  (2025)

\bibitem{shen2023advancing}
Shen, F., Ye, H., Zhang, J., Wang, C., Han, X., Yang, W.: Advancing pose-guided image synthesis with progressive conditional diffusion models. arXiv preprint arXiv:2310.06313  (2023)

\bibitem{shen2025imaggarment}
Shen, F., Yu, J., Wang, C., Jiang, X., Du, X., Tang, J.: Imaggarment-1: Fine-grained garment generation for controllable fashion design. arXiv preprint arXiv:2504.13176  (2025)

\bibitem{sudre2017generalised}
Sudre, C.H., Li, W., Vercauteren, T., Ourselin, S., Jorge~Cardoso, M.: Generalised dice overlap as a deep learning loss function for highly unbalanced segmentations. In: Deep Learning in Medical Image Analysis and Multimodal Learning for Clinical Decision Support: Third International Workshop, DLMIA 2017, and 7th International Workshop, ML-CDS 2017, Held in Conjunction with MICCAI 2017, Qu{\'e}bec City, QC, Canada, September 14, Proceedings 3. pp. 240--248. Springer (2017)

\bibitem{wang2019carafe}
Wang, J., Chen, K., Xu, R., Liu, Z., Loy, C.C., Lin, D.: Carafe: Content-aware reassembly of features. In: Proceedings of the IEEE/CVF international conference on computer vision. pp. 3007--3016 (2019)

\bibitem{wang2015training}
Wang, L., Lee, C.Y., Tu, Z., Lazebnik, S.: Training deeper convolutional networks with deep supervision. arXiv preprint arXiv:1505.02496  (2015)

\bibitem{xu2023haar}
Xu, G., Liao, W., Zhang, X., Li, C., He, X., Wu, X.: Haar wavelet downsampling: A simple but effective downsampling module for semantic segmentation. Pattern recognition  \textbf{143},  109819 (2023)

\bibitem{zhang2018road}
Zhang, Z., Liu, Q., Wang, Y.: Road extraction by deep residual u-net. IEEE Geoscience and Remote Sensing Letters  \textbf{15}(5),  749--753 (2018)

\bibitem{zhou2021nnformer}
Zhou, H.Y., Guo, J., Zhang, Y., Yu, L., Wang, L., Yu, Y.: nnformer: Interleaved transformer for volumetric segmentation. arXiv preprint arXiv:2109.03201  (2021)

\bibitem{8932614}
Zhou, Z., Siddiquee, M.M.R., Tajbakhsh, N., Liang, J.: Unet++: Redesigning skip connections to exploit multiscale features in image segmentation. IEEE Transactions on Medical Imaging  \textbf{39}(6),  1856--1867 (2020). \doi{10.1109/TMI.2019.2959609}

\end{thebibliography}

\end{document}